\documentclass{aa}
\usepackage{graphics}
\usepackage{longtable}
\usepackage{psfig}
\usepackage{astron}
\newcommand{\vmi}{(V-I)}

\begin{document}

   \thesaurus{10    
              (10.15.1;  
               03.13.6)}  

%
    \title{Luminosity and mass function of galactic open clusters: I. NGC 4815 
\thanks{
Based on observations made at the European Southern
Observatory, La Silla, Chile}.
}
    \subtitle{ }

    \author{L. Prisinzano\inst{1,2}, G.
 Carraro\inst{2}, G. Piotto\inst{2}, A.F. Seleznev\inst{3}, 
 P.B. Stetson\inst{4}, and I. Saviane\inst{5}}

    \offprints{G. Carraro}

    \mail{carraro@pd.astro.it}

    \institute{ Osservatorio Astronomico di Palermo, Piazza del
    Parlamento 1, I-90134, Palermo, Italy \and
 Dipartimento di Astronomia, Universit\`a di Padova,
Vicolo
dell'Osservatorio 5, I-35122 Padova, Italy \and Astronomical Observatory, Ural
State University, pr. Lenina 51, Ekaterinburg, 620083 Russia \and Dominion
Astrophisical Observatory, Herzberg Institute of Astrophysics, National Res
earch
Council of Canada, 5071 West Saanich Road, Victoria, British Columbia V8X 4M6,
 Canada \and UCLA/ Department of Physics and Astronomy, Los Angeles,
    CA 90095-1562}

    \date{Received xxx / Accepted xxx}

   \maketitle

   \markboth{Prisinzano et al.}{Luminosity and Mass function}

\begin{abstract} We present deep $V$ and $I$ photometry for the open
cluster NGC~4815 and four surrounding Galactic fields down to a
limiting magnitude $V\sim25$. These data are used to study cluster
spatial extension by means of star counts, and to derive the luminosity
(LF) and mass function (MF). The radius
turns out to be $3.6\pm0.3 arcmin$ at V=19.0 level, whereas the mass 
amounts at  $880\pm230 m_{\odot}$ down to V=20.8.\\ 
From the color-magnitude
diagram, we obtain the LFs in the $V$ and $I$ bands, using
both the standard histogram and an adaptive kernel. After correction
for incompleteness and field star contamination, the LFs have been
transformed into the present day mass functions (PDMF). The PDMFs from
the $V$ and $I$ photometry can be
represented as a power-law with a slope $\alpha = 3.1\pm0.3$ and
$\alpha = 2.9\pm0.3$ (the \cite{salp55} MF in this
notation has a slope $\alpha = 2.35$) respectively, in the mass range $2.5 \leq
\frac{m}{m_{\odot}} \leq 0.8$.\\ Below this mass, the MF cannot be
considered as representative of the cluster IMF,
as it is the result of the combined effect of
 strong irregularities in the stellar background,
probable internal dynamical evolution of the cluster and/or 
interaction of the cluster with the dense Galactic field.
Unresolved binaries and mass segregation can only flatten the apparent 
derived IMF, so we expect that the real IMF must be steeper 
than the quoted slope by an unknown amount.
\end{abstract}


\section{Introduction\label{intro}}
Galactic open clusters are loose and weakly bound systems, which only
in exceptional cases survive more than $10^{8}$ yrs \cite{berg80}.
Their mean lifetime is determined essentially by two factors:
the birth-place (or the orbit) and the initial mass. The dissolution of open
star clusters is mainly due to close encounters with giant molecular clouds
which are very frequent in the galactic thin disk \cite{theu92a,theu92b}.
For this reason, the age distribution is strongly biased towards young objects
\cite{wiel71}. Out of about 1,200 open clusters which are known in the
thin disk,
only 80 are older than the Hyades \cite{frie95}.\\
The dynamical evolution of open star clusters in the Galactic disk environment
has been studied by several authors
\cite{terl87,theu92a,theu92b}
To summarize, it has been shown that encounters with dense molecular
clouds are
catastrophic, whereas diffuse clouds do not completely destroy   the cluster.
Moreover, tidal effects induced by the general Galactic gravitational field
disturb significantly cluster evolution, producing distorted debris,
at the center
of which a few closely bound stars remain, with a large fraction
of binaries.\\
On the other hand, the internal dynamics of open clusters has been investigated
 by means
of N-body models \cite{aars96,fuen97},
which include the effects
of stellar evolution and a fine treatment of binaries and/or multiple
stars.\\
Among the others, a very important result is that
the cluster evolution crucially depends on the Initial Mass Function (IMF),
and on the cluster richness.
To go in some detail, de la Fuente Marcos \cite*{fuen97} finds that
power-law \cite{salp55} like IMFs accelerate
the cluster disruption if the initial population is small, and that at the
end of the evolution a huge number of binary systems with the same mass is
formed.\\

An important constraint for all these models would be the possibility to
derive the IMF
from observations of open star clusters.
Unfortunately, this is a very difficult task for several reasons.\\
First, most open clusters are poorly populated
ensembles of star, containing from tens to hundreds
of stars, and only in rare cases up to some thousands.
Secondly, they are located in the inner regions of the thin disk. This fact
renders  it
very difficult to detect cluster members
(at odds with globular clusters), due to the high contamination of the field
stars in the same line of sight
as the clusters. Third, they can be strongly obscured due
to  interstellar
absorption lying between us and the cluster \cite{jane82}.

Probably, the most challenging task is to obtain cluster membership.
This is feasible for objects close to the Sun, for which 
we can obtain high quality radial velocity and
proper motion measurements. Going farther away
one has to rely on statistical corrections which usually are made by
comparing the cluster with the field in its outskirts.  Ideally, when
a correction for field star contamination has been performed, a
reasonable actual (Present Day) mass function (PDMF) can be derived
from the observed LF.  
By modeling the dynamical evolution of the cluster the PDMF can be
converted into an IMF which can be compared with the MFs of other
clusters to look for universality or deviations.

In this paper, the first of a series dedicated to obtaining luminosity
and mass functions for Galactic open clusters, we present results for
NGC~4815, a rather rich, distant open cluster located at about 2.5 kpc
from the Sun .\\ 
The ultimate goal is to
provide a sample of PDMFs as a function of cluster age, population,
and location to serve as template for N-body models.\\ 
We plan to study other clusters with the same 
method in forthcoming papers.
In order to obtain a good representation of the field
stellar population, in addition to the cluster we observed four nearby
Galactic fields, whose population has been averaged and used for the
statistical subtraction of the foreground/background objects
from the cluster.  An adaptive kernel estimator was then applied to
obtain a continuous LF, free of binning bias,  
together with the standard histogram
technique. Finally, the MF has been derived from $m = 4 m_{\odot}$ down
to $m = 0.35 m_{\odot}$ by using the luminosity-mass relation from the
Girardi et al. \cite*{gira00} models.

\begin{table} [htb]
\tabcolsep 0.7truecm
\caption {NGC 4815 fundamental parameters.Age, distance
and reddening are taken from the study of Carraro and Ortolani (1994)}
\begin{tabular}{ll} 
\hline
\hline
$l$ & 303.63\\ 
$b$ & -2.09 \\ 
$E(B-V)$ & 0.70$\pm$0.05 \\
$(m-M)_V$& 14.10$\pm$0.05 \\
$(m-M)_I$& 13.20$\pm$0.05 \\
Age &($0.5\pm0.25) \times 10^{9} yr$\\ 
$Z$&$\approx$0.008\\
\hline
\hline
\end{tabular}
\end{table}

The layout of the paper is as follows. 
Section~2 summarizes our present knowledge of the open cluster NGC~4815.
In Section~3 we present the
observations and data reduction technique; Section~4 is dedicated to
the discussion of the CMDs obtained for the cluster and the
field. Section ~5 is dedicated to a detailed analysis of the star counts,
while Section~6
deals with the LFs derived with the two
different techniques. Section~7 presents the MF of NGC~4815 and the
comparison with other MFs. Finally Section~8 summarizes our results.


\section{Previous investigations}
NGC~4815 has been studied in the past by several authors.\\ The first
photometric data have been collected by Moffat \& Vogt \cite*{moff73},
who studied 9 stars in the cluster region, but did not recognize any
sequence.\\ 
Kjeldsen \& Frandsen \cite*{kjel91} presented a catalogue
of UBV photometry for 559 stars.\\
Carraro and Ortolani \cite*{carr94} firstly made an accurate analysis
of the CMD by obtaining BV photometry of 2500 stars down to V=19.5.
Their results led to the determination of the cluster parameters:
NGC~4815 turns out to be of intermediate age (about half a Gyr old,
like the Hyades), with a reddening $E(B-V) \approx 0.70$, and a
distance of about 2.5 kpc from the Sun.  The metallicity is
half Solar (see Table~1).  
From a preliminary analysis of the MS mass
function they obtained an index $\alpha=2.55$, slightly higher than
the classical Salpeter value.\\ These data have been used to study the
cluster structure and luminosity function with a different method by
Chen et al. \cite*{chen98}, who obtained for the first time the
cluster mass function by subtracting the field contamination with a
Galaxy structure model.  In the mass range $1.1 < \frac{m}{m_{\odot}}
< 2.2$ they obtain a slope
$\alpha=1.97\pm0.17$, 
and a total mass of 912
$m_{\odot}$ to the photometric limit (V = 19). They finally found
evidence for mass segregation.

\section{Observations and reduction\label{obs}}


\subsection{Observations  \label{data}}
%
NGC~4815 has been observed with the ESO-NTT telescope on April 14-16,
1999. We used the EMMI red arm, equipped with the 2048$\times$2048
ESO$\#36$ Tektronics CCD.  The scale on the sky is  0.27
arcsec/pixel, for a total field of $9.2\times9.2$ arcmin$^{2}$. The
observing log is in Table~1. Almost all the images have been collected in
sub-arcsec seeing conditions and in photometric conditions.  During
the same nights, four Galactic fields have been observed. These fields
are located at 30 arcmin from the center of NGC~4815, on the North,
South, East and West sides, respectively. The exposure times and the
seeing of the Galactic field images are the same as for the cluster
data, with the only exception of the southern one, for which only
short exposures could be obtained. We opted to exclude
this field in the statistical analysis of luminosity and mass function.


\begin{table} [htb]
\tabcolsep 0.5truecm
\caption {Log-book of the NGC 4815 observations}
\begin{tabular}{cccc} 
\multicolumn{1}{c}{Field} &
\multicolumn{1}{c}{Filter}&  
\multicolumn{1}{c}{Exp. Time)} &
\multicolumn{1}{c}{Seeing} \\  
& & [s] & FWHM[$^{\prime\prime}$] \\
\hline
 Center & {\it V} &   15 & 0.8 \\
 Center & {\it I} &   15 & 0.8 \\
 Center & {\it V} &  900 & 0.8 \\
 Center & {\it I} &  900 & 0.8 \\
 West   & {\it V} &   15 & 1.0\\
 West   & {\it I} &   15 & 1.1 \\
 West   & {\it V} &  900 & 1.0 \\
 West   & {\it I} &  900 & 0.9 \\
 North  & {\it V} &   15 & 0.7 \\
 North  & {\it I} &   15 & 0.8 \\
 North  & {\it V} &  900 & 1.0 \\
 North  & {\it I} &  900 & 0.9 \\
 East   & {\it V} &   15 & 0.7\\
 East   & {\it I} &   15 & 0.7 \\
 East   & {\it V} &  900 & 0.9 \\
 East   & {\it I} &  900 & 0.9 \\
 South  & {\it V} &   15 & 0.8 \\
 South  & {\it I} &   15 & 0.6 \\
\hline\\
\end{tabular}
\end{table}

\begin{figure}
\centerline{\psfig{file=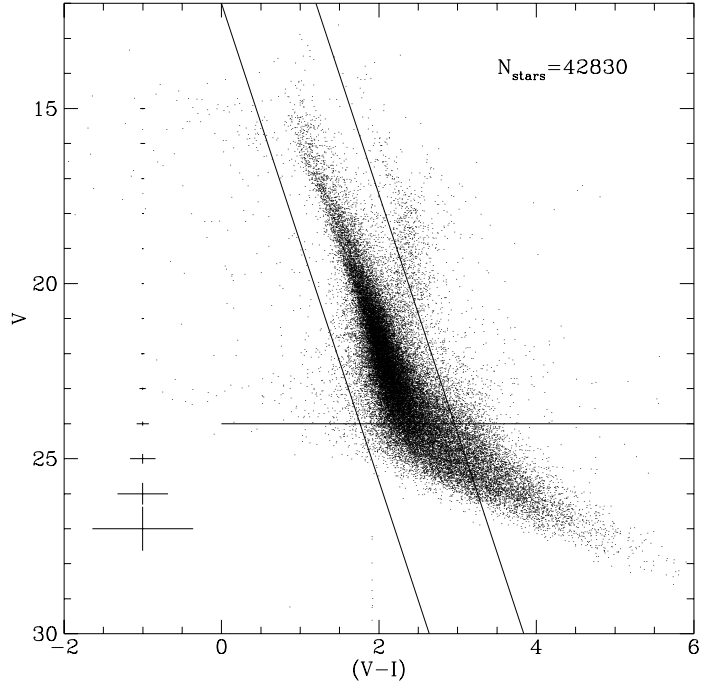,width=10cm,height=10cm}} 
\caption{The CMD of NGC 4815. A selection based on the
sharp and chi  parameters has been done following Stetson (1987). 
Sharp is vaguely related to the intrinsic (i. e. outside the atmosphere)
angular size of the astronomical object. For an isolated star, sharp
should have a value close to zero, whereas for semi-resolved galaxies
and unrecognized blended doubles 
sharp will be significantly greater than zero, and for cosmic rays and
some image defects will be significantly less than zero.
Chi is a robust estimate of the observed pixel-to-pixel
scatter from the model image profile divided by the expected
pixel-to-pixel scatter from the image profile.
The straight lines indicate the CMD region used to derive the LFs and
represents the limiting magnitude for the $50\%$ completeness level.
 Finally, the crosses on the left side of the plot represent the 
photometric errors at that magnitude level.}
\label{cal}
\end{figure}

\subsection{Data reduction and calibration}
All the images have been pre-processed in the standard way with IRAF, using
the sets of bias and sky flat-field images collected during the observing
nights.
The instrumental magnitudes, and the  positions of the  stars for each frame
have been derived by profile-fitting photometry with the  package DAOPHOT II
and ALLSTAR \cite{stet87}.
Then we used ALLFRAME \cite{stet94} to obtain the simultaneous
PSF-fitting photometry of all the individual frames.
In order to obtain the transformation equations relating the
instrumental ($vi$) magnitudes to the standard {\it V\/} (Johnson),
$I$(Kron-Cousins) system we followed the procedure already described
in \cite{Bed20}. Seven Landolt (1982) fields of
standards have been observed. Specifically: the PG0918+029,
PG0942-029, PG1047+003, PG1525-071, PG1657+078, SA\-104\-334,
and SA\-110-\-496 regions. In each of these, there are other secondary
standard stars by Stetson \cite*{stet00}, which extend the previous
Landolt sequence.  For all of these stars, aperture photometry was
obtained on all the images after the digital subtraction of
neighboring objects from the frames.  We used
DAOGROW \cite{stet90} to obtain the aperture growth curve for each
frame and other related codes (by PBS) to determine the aperture
corrections, the transformation coefficients, and the final calibrated
photometry ({\it e.g.}, Stetson 1993).  We used transformation
equations of the form:

$$ v = V + A_0 + A_1 * X + A_2 * \vmi,$$ 
$$ i = I + B_0 + B_1 * X + B_2 * \vmi,$$ 


\begin{table} [htb]
\tabcolsep 0.5truecm
\caption {Coefficients of the calibration equations}
\begin{tabular}{ccc} 
\hline
&April 14,1999 & April 15,1999 \\
\hline
$A_0$ &0.053$\pm$0.007  &0.044$\pm$0.006\\
$A_1$ &0.249$\pm$0.041  &0.259$\pm$0.023\\
$A_2$ &-0.036$\pm$0.002 &-0.028$\pm$0.003\\
$B_0$ &0.690$\pm$0.012  &0.601$\pm$0.012\\
$B_1$ &-0.054$\pm$0.061 &0.307$\pm$0.050\\
$B_2$ &0.020$\pm$0.002  &0.030$\pm$0.005\\
\hline
\end{tabular}
\end{table}

\noindent
where the values of the coefficients are listed in Table~3
for the two nights.
In these equations $v$ and $i$ are the aperture magnitude already
normalized to 1 sec, and $X$ is the airmass.
Second order color terms were tried and turned out to be negligible in
comparison to their uncertainties. It is a reasonable hypothesis that
the color-dependent terms imposed by the telescope, filter, and
detector should be reasonably constant over the course of a few
nights, so after $A_2$ and $B_2$ had been independently determined
from each night's data, we obtained weighted mean values of these
parameters and imposed them as fixed quantities for the definitive
calibration. The NGC$\,$4815 images enabled us to define local
standards that met the following conditions: each was well separated
from its neighbors (according to a criterion described in Stetson
\cite*{stet93} \S4.1), each was observed in all frames, each had a
mean value of the goodness-of-fit statistic, CHI, less than 1.5. Once
calibrated, this local standard sequence enables us to place all of
observations of NGC~4815, and of the Galactic fields 
on the same magnitude system with an uncertainty less than 0.005
magnitudes (internal error).

\subsection{Artificial star tests}
In order to obtain the LF, we estimated the completeness of our
samples. The completeness corrections have been determined by standard
artificial-star experiments on our data. For each field, we performed
five independent experiments adding on the original frames as large  a
number as
possible of artificial stars in a spatial grid as in Piotto
\& Zoccali \cite*{piot99}. We performed the photometry of the frames
with the added artificial stars following all the steps and using the
same parameters as for the original images. We converted the new
instrumental magnitudes to the standard photometric system using the
coefficients of the calibration equations of Section~2.2.

In the LFs we include only the values for which the completeness
corrections, defined as the ratio between the number of found
artificial stars to that of the original added ones, were 50\% or
higher. This sets the limiting magnitude to V = 24.
       
\begin{figure*}
\centerline{\psfig{file=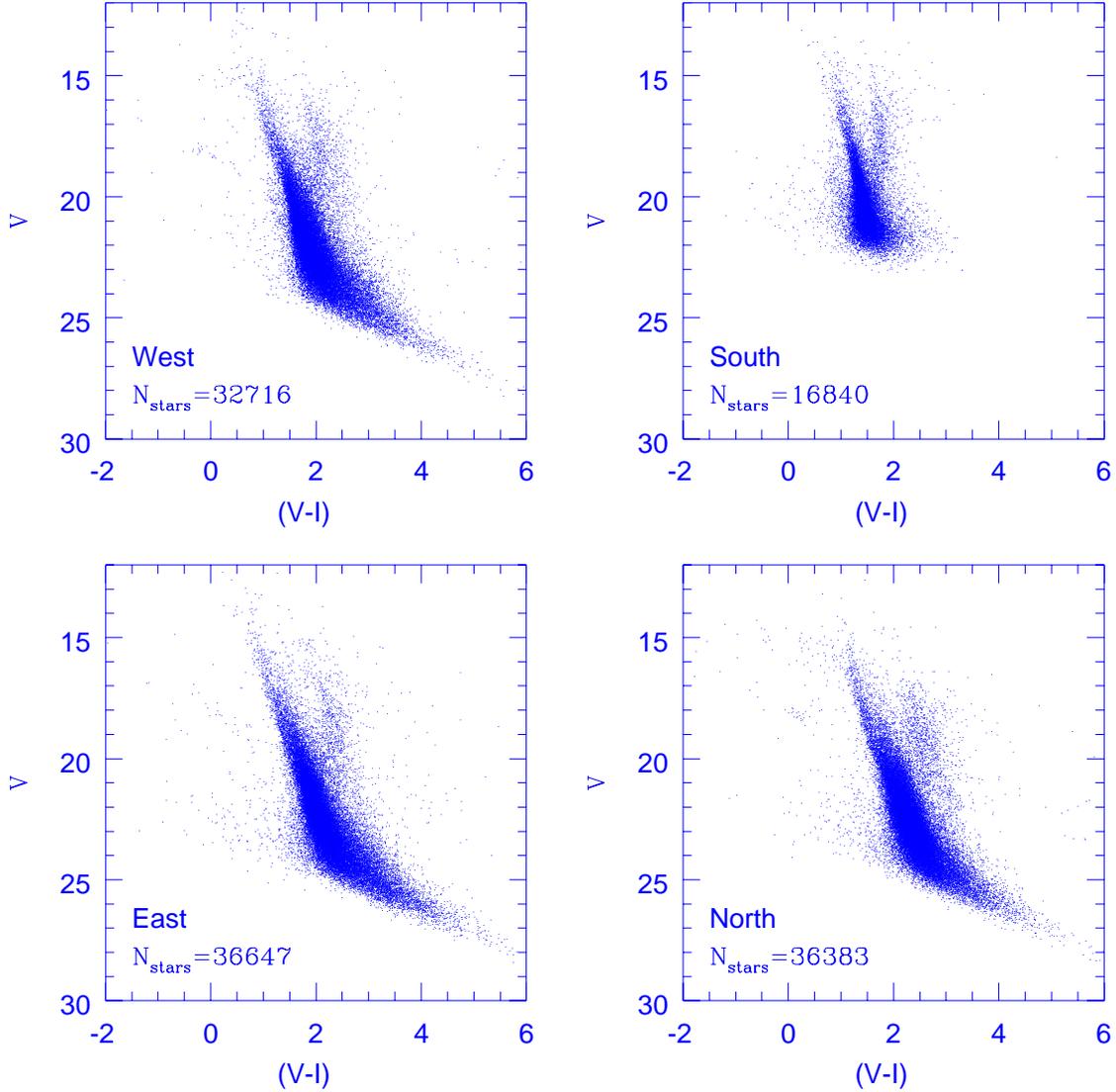,width=16cm,height=16cm}}
\caption{ CMDs of the observed nearby fields. Each field is centered
at $30'$ from the center of NGC~4815. The southern field do not have
deep exposures, and has not been used in the paper.}
\label{east}
\end{figure*}

\section{The Color-Magnitude Diagrams\label{cmd}}
We obtained the V and I photometry with the aim of determining the LF
(and MF) of the main sequence stars in NGC~4815. We
used two colors as we wanted also to derive the CMD, which allows to:
{\it i)} discriminate stars from false detections; {\it ii)} better
identify the cluster population, in particular the MS stars.\\ The CMD
for NGC~4815 is shown in Fig. \ref{cal}, whereas the CMDs for the four
nearby fields are presented in Fig. \ref{east}.  Clearly, the CMD of
the cluster field does not differ much from those of the surrounding
Galactic disk. This means that NGC~4815 is actually badly contaminated
by background/foreground stars, as expected from its particular
position in the Galactic thin disk.  The cluster field (Fig.~1) is much
more populated than the neighboring fields (Fig.~2), confirming that NGC~4815
is a real open cluster (see also the discussion in Section~5).
\noindent
Our photometry reaches $V \approx25$ at the base of the MS.\\ 
The MS extends almost vertically for more than 10 magnitudes,
from V = 15 to V=25, although
the completeness analysis prevents us from using stars dimmer than
V= 24. 
The width of the main sequence gets greater at larger
magnitudes. This is partly  due to the increasing  photometric errors
at increasing magnitudes (see Fig.~1).
However the MS is much wider than expected simply from
photometric errors.  
Between the various causes which concur to 
enlarging the natural MS width we can envisage the presence
of unresolved binary stars, common in open clusters, and a possible
spread in metals.
The red vertical sequence, which detaches from the MS at about
V = 22.5 represents the giant branch of the Galactic disk
population.\\
At odd with the previous
photometry, the present data (see Fig.~1)
do not allow us to recognize clearly the
presence of a red giant clump, better visible in the BV photometry of
Carraro and Ortolani \cite*{carr94} at $V \approx 14.3$.  
This is due to the saturation of
many bright stars above V = 14.5, which were not of interest for the
purposes 
of
this work.

\section{Star counts and  cluster size}
The aim of this section is to obtain the surface density distribution
of NGC 4815, and derive the cluster size and extension
in the magnitude space  by means of star counts. The cluster radius is
indeed one of the most important cluster parameters, useful 
(together with cluster mass) for the 
determination of cluster dynamical parameters. 
Star counts allow to determine statistical properties
of clusters (as visible star condensations) with respect to the
surrounding stellar background.

Star counts have
been performed following the methods described in \cite{dan86},
\cite{sel94} and \cite{sele98} by evaluating the 
functions $N(r)$ and $N(V)$ ($N(I)$).
$N(r)$ represents  star number in the circle of
radius $r$ and $N(V)$ is star number up to apparent magnitude $V$, i.e.
integral luminosity function. In order to obtain $N(r)$ we divided the
cluster field (and comparing fields) in concentric rings 40 pixels
(11 arcsec) wide and counted the number of stars
falling inside each ring. Then $N(r)$ is obtained by summing the  corresponding
ring star numbers. Similarly, in order to obtain $N(V)$, we divided stars
in the various fields in magnitude bins $0.2^m$ wide and summed up corresponding
bin star numbers. Star counts with $N(r)$ give us cluster radius and
cluster star number, star counts with $N(V)$ give us the magnitude
of faintest cluster stars (up to the limiting magnitude).

Due to the saturation of the brightest stars ($V<14.5$), some regions
of the cluster cannot be used to perform star counts. In these cases,
for $N(r)$ determination we used the stars located in the same regions
from the work of Carraro \& Ortolani (1994).

A smoothed surface density $F(r)$ is then obtained by differentiating
numerically the best fit least squared polynomial approximation
for $N(r)$  according to the equation:

\begin{equation}
\label{differ}
N(r)=2\pi \int_0^r r^\prime F(r^\prime) dr^\prime
\end{equation}

\noindent
(see Seleznev (1994), for details). The results are shown in Fig.3, where
the logarithm of the surface density profile is plotted as a function of
the limiting magnitude. From this figure it is evident that the cluster
significantly emerges from the background only for $V$ less than 21.
Below this value  NGC 4815 is completely confused with the Galactic field.
Moreover we note  a significant  density decrease at the cluster position.

Integral distribution functions are then obtained both for cluster field and
for comparing fields. Let us consider below a theoretical  integral
distribution function $N_i(x)$, where $x$ can be $r$, $V$ and $I$, and 
$i=0$ indicates cluster field, while $i=1,2,3,...$ indicate comparing
fields.

Following \cite{dan86} and \cite{sele98} we 
consider the normalized deviation $z$ of $N_i(x)$ from the
corresponding average number of field stars:

\begin{equation}
\label{z}
z_i(x) = \frac{\left| N_i(x)-\overline{N(x)} \right|}{\sigma (x)}
\hspace{1cm} (i=0,1,\ldots,k),
\end{equation}

\noindent
where

\begin{equation}
\label{mean}
\overline{N(x)}=\frac{1}{k} \cdot \sum_{i=1}^k N_i(x)  ,
\end{equation}

\noindent
and

\begin{equation}
\label{disp}
\sigma^2(x)=\frac{1}{k-1} \cdot \sum_{i=1}^k \left[ N_i(x)-
\overline{N(x)} \, \right]^2 .
\end{equation}

\noindent
The normalized deviations $z_i$ are connected with the probability that the i-th
field significantly deviates from other fields. Then $z$ values are
usually  called "significance levels".

There are two criteria to determine cluster boundaries in $x$-space.
The first criterion is as follows. One has to look for a
value $x_c$ such that
$(N_0(x)-\overline{N(x)})=N_c=const$ for all $x\ge x_c$ and
$z_0(x)\ge z_i(x)$ ($i=1,2,3,...$) for all $x\le x_c$.
In this case we can conclude  that the cluster has the boundary at the point
$x=x_c$ and $N_c$ is cluster star number, and we can estimate an error
for $x_c$ and $N_c$ (Danilov et al. 1986).

The second criterion is as follows. The cluster has no boundary in the
sense of the first criterion, but for $x>x_{c2}$ $z_0(x)< z_i(x)$ for
some $i$, i.e. i-th field deviates from mean $\overline{N(x)}$ more
significantly than cluster field. This situation means that we can
determine only a lower
estimate of the cluster boundary, i.e. $x_c> x_{c2}$ and $N_c> N(x_{c2})$.
The reason is related to large-scale fluctuations of integral distribution
function in the comparing fields. Unfortunately this situation 
frequently occurs for open star clusters located in a
strongly fluctuating background.

The star counts in $V$ and $I$ spaces have been performed both for central part of
NGC 4815 (circle with $r\approx 2\hspace{0.2cm} arcmin$ in the central
field, see Table 2) with twelve
comparing fields (four circles in each East, West and North fields)
and for all cluster field (i.e. the entire central field)
with three comparing fields (East, West and North).

The star counts for the central part of the cluster confirm that it is less
populated by faint stars ($V>18^m-20^m$) than the other fields (see Fig.3).
We obtained cluster boundary $V_c=20.8^m\pm 0.2^m$ following the 
first criterion (see above). The trend of $z_0(V)$ is shown in
Fig.4, where it is compared with mean $z$ for North, West and East
fields. The luminosity function of cluster
stars (see following section), that is differential distribution
function, has negative value for $V>V_c$.
Presumably, the reason is the higher light absorption along line
of sight beyond the central part of the cluster.

\begin{figure}
\centerline{\psfig{file=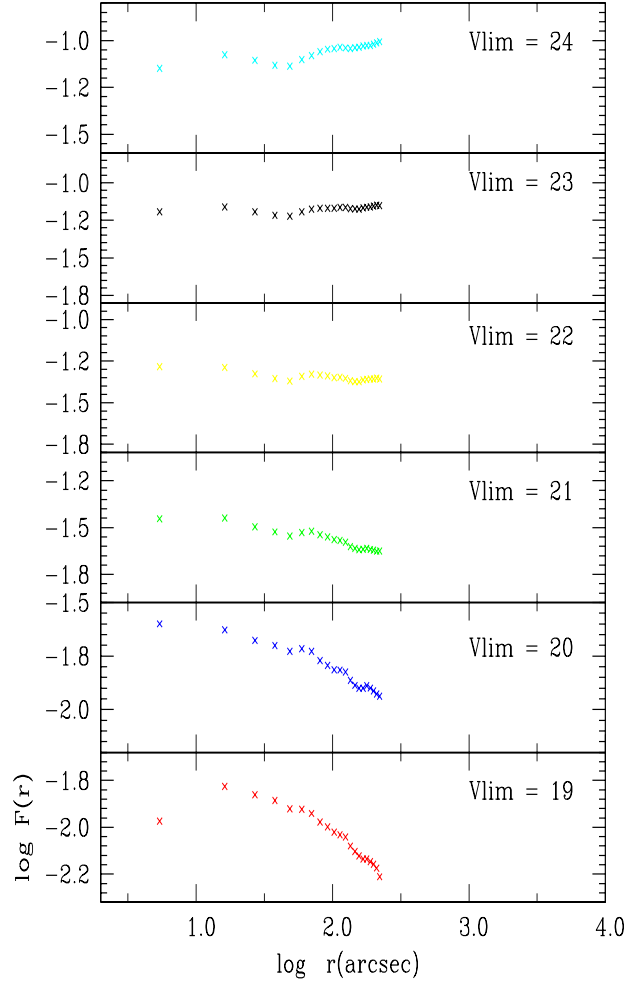,width=10cm,height=14cm}}
\caption{Radial surface density profiles of NGC~4815 for 6 different limiting
magnitudes. Surface density is in stars per arcsec$^{2}$.}
\label{pro}
\end{figure}

\begin{figure}
\centerline{\psfig{file=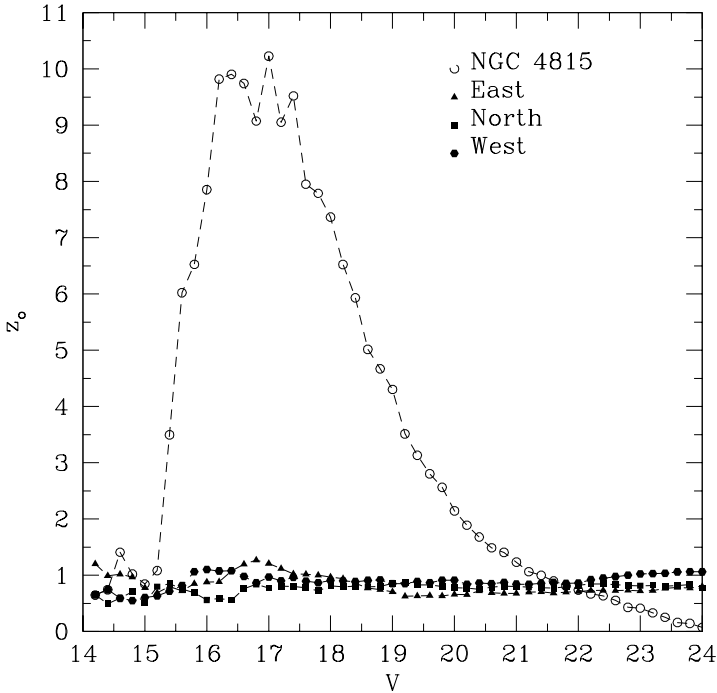,width=10cm,height=10cm}}
\caption{Significance level for cluster and surrounding fields.
See text for details.}
\label{rela}
\end{figure}

By performing star counts in $V$ and $I$ for all the central field we did not
obtain the cluster boundary according to first criterion. In the $I$ passband
$N_0(I)$ significantly deviates from mean $\overline{N(I)}$
up to $I\approx 22^m$, but in the $V$
$N_0(V)$ significantly deviates from mean $\overline{N(V)}$
only up to $V\approx 20^m$. However 
$(N_0(V)-\overline{N(V)})$ and $(N_0(I)-\overline{N(I)})$ 
increase for much fainter magnitudes, and the corresponding
luminosity functions are positive (see Fig.5 and 6 in the following
section). 
This contradictory result is due to the  very complicate
character of the stellar background in the vicinity of NGC~4815 and
beyond it. It means that mean integral distribution function
$\overline{N(V)}$ in comparing fields increases more slowly than
$N_0(V)$,  but the dispersion $\sigma(V)$ is too large.

Because the cluster center is slightly shifted with respect to  the center of
the central field, star counts in $r$-space can be carried out in the entire
circle only to radius $r=880~pixel$ (4~${\it arcmin}$). 
Due to this reason
star counts were made also in sector of $199^\circ$ to radius $r=1080~pixel$
(4.9~${\it arcmin}$). The results of star counts for different limiting
magnitudes are listed in Table~4.

\tabcolsep=0.8cm
\setcounter{table}{3}
\begin{table*}[t]
\caption{Results of star counts for cluster size and star number
determination}
\medskip
\begin{tabular}{ccccc}
\hline
$V_{lim}$ & \multicolumn{2}{c}{Entire circle ($r_{max}=880~pixel$)} &
    \multicolumn{2}{c}{Sector $199^\circ$ ($r_{max}=1080~pixel$)}\\
\cline{2-5}
        &   $R_c$     &   $N_c$   &  $R_c$      &  $N_c$    \\
\hline
$16^m$  & $800\pm 71$ & $89\pm 6$ & $800\pm 55$ & $89\pm 2$ \\
\hline
$17^m$  & $800\pm 79$ & $219\pm 13$ & $800\pm 40$ & $261\pm 3$ \\
\hline
$18^m$  & $800\pm 56$ & $361\pm 11$ & $840\pm 56$ & $454\pm 5$ \\
\hline
$19^m$  & $800\pm 64$ & $447\pm 22$ & $>1080$ & $>637$ \\
\hline
$20^m$  & $>680$ & $>319$ & $>1080$ & $>599$ \\
\hline
$21^m$  &  \multicolumn{4}{l}{Cluster does not deviate from comparing
                            fields significantly, $N_c>0$} \\
\hline
$22^m$  &  \multicolumn{4}{l}{Cluster does not deviate from comparing
                            fields significantly, $N_c<0$} \\
\hline
$23^m$  &  \multicolumn{4}{l}{Cluster does not deviate from comparing
                            fields significantly, $N_c<0$} \\
\hline
$24^m$  &  \multicolumn{4}{p{12cm}}{Cluster significantly deviates from
                comparing fields as minimum, $N_c<0$} \\
\hline
$24.7^m$  &  \multicolumn{4}{p{12cm}}{Cluster significantly deviates from
                 comparing fields as minimum, $N_c<0$.
For $r\in[800,880]$ $N_c>0$ and for $r\in[840,880]$ cluster significantly
deviates from comparing fields as maximum.} \\
\hline
\end{tabular}
\end{table*}

\noindent
First column contains limiting magnitude, second and third columns
contain cluster radius and cluster star number from counts in the entire circle,
fourth and fifth columns contain the same values from counts in
$199^\circ$ sector. Star number in fifth column is obtained as result of
star counts multiplied by (360/199).

The comparison of results of star counts in the entire circle and in the sector shows
that the asymmetry of the cluster increases at increasing limiting magnitude.
Beginning from $V_{lim}=21^m$ the cluster disappears as a statistically significant
stellar density fluctuation, and below $V_{lim}=24^m$ the cluster region appears
as stellar density minimum - statistically significant comparing with three
fields. The most probable reason is the presence of absorbing cloud just beyond
the cluster.

Star counts with stars up to $V_{lim}=19^m$ gave us
the estimate of cluster radius, which turns out to be $3.6\pm 0.3$ {\it arcmin},
significantly lower than $4.6$~${\it arcmin}$ estimated by Carraro \& Ortolani
(1994). However, result of Carraro \& Ortolani (1994) can be regarded only
as preliminary, because cluster radius was determined only by surface density
profile as distance to a point where the slope of this curve changes, being the
slope negative along all the profile.
It is difficult to compare our results with previous results
directly, because now we are using a quantitative statistical method. Unfortunately,
the cluster radius of $3.6$~${\it arcmin}$ is very close to the limit of
investigated field, and we can not regard this result as definitive.
Furthermore, $3.6$~${\it arcmin}$ corresponds to a radius of $2.6~ pc$ at the distance
of $2.5~ kpc$. This value lies in the lower limit of cluster radii interval
according to \cite{dan94}. Then it is quite possible
that we have determined the radius of the cluster core.

It is necessary to note that the central field contains much more stars
than any comparing field (see Fig.1 and Fig.2). This excess in star
number is mainly due to the  periphery of the central field which is very rich
in faint stars. Unfortunately our star counts can not unambiguously answer the 
question -
is the cluster really disappearing at $V_{lim}=21^m$ and has radius
of about 3.6 arcmin or faint stars in the outskirts of central field belong
to cluster and form its halo ?

Firstly, star counts in wider fields are necessary (both for cluster
and comparing fields).  Desirable radius of investigated field
is of about 10-15 pc (i.e. 14-21 arcmin). Secondly, it is necessary
to study color extinction for stars of different magnitudes, i.e. to
observe in three colors.\\
\noindent
On the basis of the previous discussion, we are going to estimate 
the luminosity function
for all the central field in the magnitude range $V\in[14^m;24^m]$.

\section{Luminosity Function\label{lfs}}
The luminosity function of NGC~4815 was obtained as follows.
In the CMDs, two lines, to the left and right of the main sequence,
defined by the two relations $V=6.8(V-I)+12$ and $V=6.8(V-I)+3.84$, were
used in order to select only the main sequence stars (Fig. 1). For each field,
an adaptive kernel estimator was then applied to obtain a continuous LF.
The LF was also derived by the standard histogram technique. A detailed
description of the application of the kernel function method for estimating
distribution functions can be found in \cite{silv86}, and in \cite{sele98}
and \cite{sele00}.

The advantage of this method is twofold: 1) first of all it is not necessary
to define a bin, whose size usually is arbitrary and does not allow to detect
unambiguously all the statistically significant features present in the LF;
the estimate of the LF by a kernel method with its confidence interval is
an estimate of an unknown density of a stellar population, while the histogram
alone gives only the distribution of a specific sample; 2) the form of a
histogram - crucial for the determination of the LF slope - depends both
on the bin width and on the initial bin location. The distribution function
provided by the kernel estimate only depends on the kernel width. Moreover
criteria have been developed to fix the kernel width with high degree of
confidence \cite{silv86}.

\begin{figure}[!h]
\centerline{\psfig{file=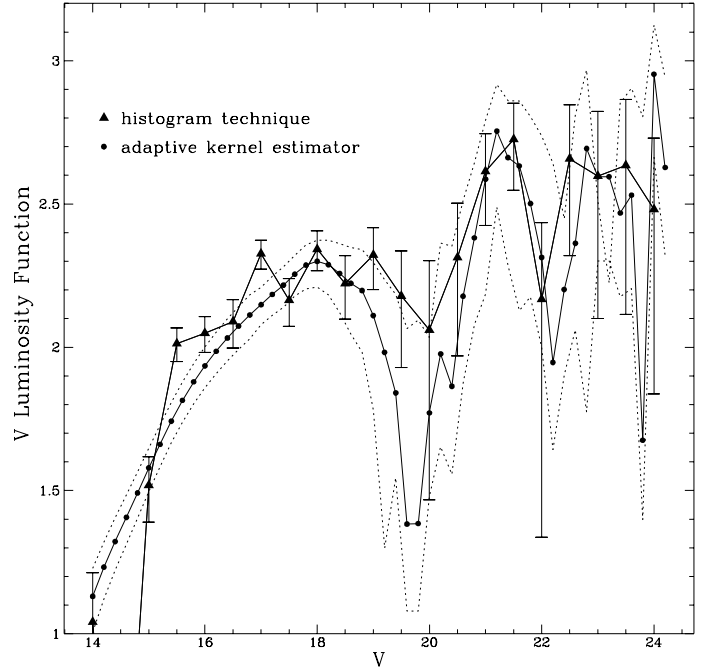,width=10cm,height=10cm}}
\caption{The $V$ LFs obtained with both the standard histogram
technique ({\it triangles}) and with an adaptive kernel estimator
({\it circles}). The {\it dotted curves} are the confidence intervals for the
adaptive kernel estimator. Note the large error bars below $V=21$,
when the clusters starts to vanish into the Galactic field.}
\label{flv}
\end{figure}

\begin{figure}[!h]
\centerline{\psfig{file=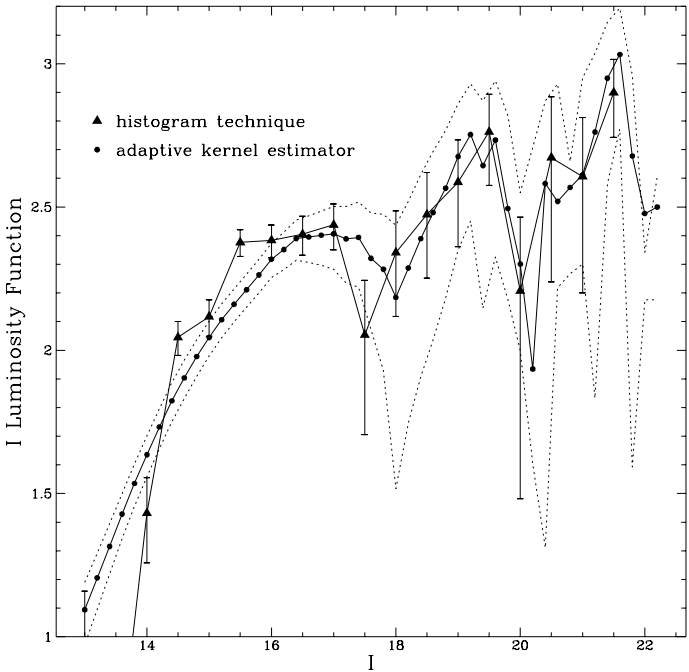,width=10cm,height=10cm}}
\caption{As for Fig.~5, but for the I filter.}
\label{fli}
\end{figure}

The adaptive kernel estimate of the density distribution in one-dimensional
case is calculated by the formula:

\begin{equation}
\label{lf}
\hat f(x)=\frac{1}{n} \sum_{i=1}^{n} h \lambda_i \cdot
K\left( \frac{x-X_i}{h\lambda _i} \right),
\end{equation}
where n is the size of the observing sample $X_1...X_n$ of the random variable
$x$, $K$ is the kernel function which must satisfy the condition:

\begin{equation}
\label{k=1}
\int \limits_{-\infty}^{+\infty} K(x)dx=1 ,
\end{equation}
and $\lambda_i$ are the local window width factors. The adopted kernel has
the form

\begin{equation}
\label{k}
K(x)=\frac{3}{4\sqrt{5}} \left( 1-\frac{x^2}{5} \right)
\end{equation}
when $|x|\leq \sqrt{5}$, and zero elsewhere.

Basically eq.\ref{lf} spreads each observed value $X_i$ of the random variable
$x$ over an interval (window) centered on $X_i$ with the probability density
$K\left( \frac{x-X_i}{h\lambda_i} \right)$. The kernel width $h\lambda_i$
can differ from point to point. The advantage of having an adaptive kernel
is that it is easy to properly sample the tails of the distribution, just
playing with the kernel width, which is set larger in low-density regions
automatically (see \cite{silv86}). Parameter $h$ was set to be $0.3^m$.

To determine the confidence interval for the resulting LF, we used the method
described in Seleznev et al. (2000) and references therein.

For each field, the LFs have been corrected for incompleteness. The LFs of the
external fields were used to subtract the field-star contamination from the
LF of the central field.

To this end we defined $\hat f_i(V)$ as the density of the distribution of
the stellar $V$ magnitudes, where $i$ is the number of the field (here we
assume that $i=0$ corresponds to the cluster field or central field and
$i=1,2,3$ correspond to the external fields). The luminosity function of
the cluster stars is given by

\begin{equation}
\label{clusterLF}
\hat f(V)=\hat f_0(V)-\frac{1}{k} \sum_{i=1}^{k} \hat f_i(V).
\end{equation}

The same procedure was repeated to get the $I$ LF. The $V$ and $I$ LFs obtained
both with the histogram and the adaptive kernel techniques for all the central
field are shown in Figs.5 and 6. Note the large uncertainties below $V=21$
($I\sim 19$), where fluctuations of stellar background density play a 
significant role.
LFs obtained by different methods are close each other due to large volume
of the sample. Note that the stellar content below $V=20$ ($I=18$) is due to 
outskirts of the central field (see previous section).

\begin{figure}[!h]
\centerline{\psfig{file=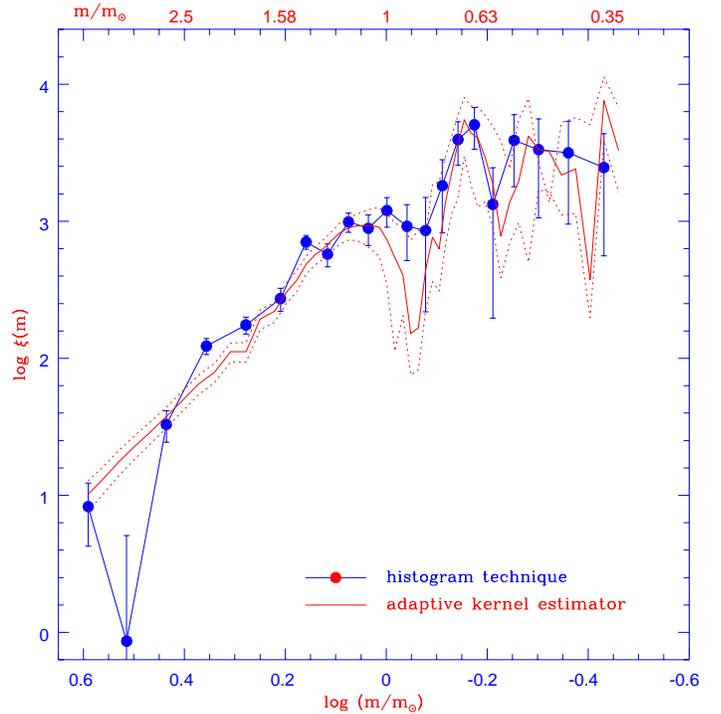,width=10cm,height=10cm}}
\caption{Comparison of the MFs of NGC~4815 from $V$ LFs derived by the
adaptive kernel estimator and the standard histogram technique. The
{\it dotted curves} are the confidence intervals for the adaptive
kernel estimator.}
\label{fmv}
\end{figure}

\begin{figure}[!h]
\centerline{\psfig{file=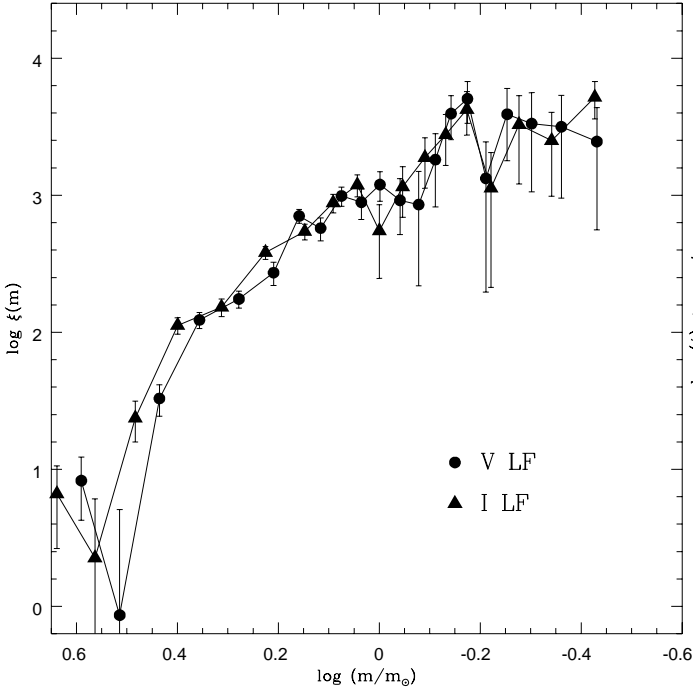,width=10cm,height=10cm}}
\caption{Comparison of the MFs of NGC 4815 from the $V$ and $I$ LFs 
derived by the standard histogram technique.}
\label{fmvi}
\end{figure}

\begin{figure}[!h]
\centerline{\psfig{file=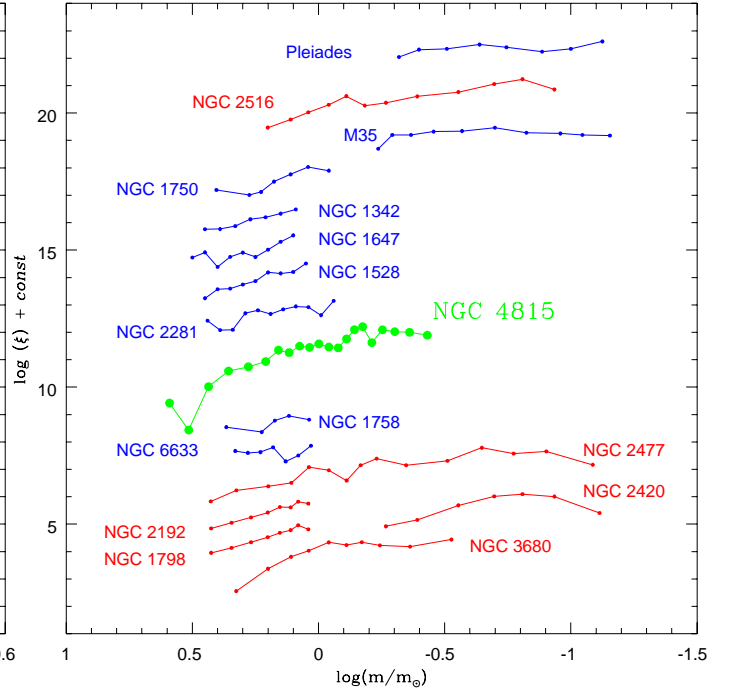,width=10cm,height=10cm}}
\caption{Comparison of the NGC 4815 MF with other open cluster MFs.
For display purposes the MFs are shifted by an arbitrary constant.}
\label{allfm}
\end{figure}

\section{Mass Function\label{mfs}}
The LFs can be transformed into MFs using a mass-luminosity relation (MLR).
Since we cannot obtain an empirical transformation, we must rely on the
theoretical models. Therefore, we used the ZAMS isochrones by Girardi et
al. (2000) with a metallicity Z=0.008. The distance modulus and the reddening
that result from the fit with the model are in agreement, within the errors,
with the values found by Carraro and Ortolani (1994). We adopted $E(B-V)=0.70$
and $(m-M)_V=14.10$. From these values we get $(m-M)_I=13.20$ (see Table
1). Fig. 7 shows the MF from the V LFs obtained by the adaptive kernel
estimator and the standard histogram technique. The two MFs are rather similar.
There is also a substantial agreement between the MFs from the V and I LFs
(Fig. 8), showing an internal consistency of both the models and the method.\\

By looking at Fig.~7, it appears that  there is a dip in 
MF at $m\simeq 0.9M_{\cdot}$.
This dip in MF is a the consequence of the  corresponding dips in LFs at
$V\simeq 19.5^m-20^m$ and $I\simeq 17.5^m-18^m$ (see Fig.s 5 and 6).
To estimate the statistical signifincance of this dip, we concentrate on
"Adaptive kernel estimator" method, which is
is more sensitive to peculiar properties of LF
and MF than the histogram technique (see previous sections).\\
We can use the ratio $\eta$ of the depth of this minimum (relatively to
the mean smooth MF level) to the dispersion of MF estimate in the
vicinity of MF minimum as a measure of 
the statistical significance of the MF minimum.
As for the MF minimum at $m\simeq 0.9M_{\cdot}$ we obtain $\eta\simeq 1.98$ from
the "adaptive kernel" MF estimate, which  corresponds to a reality probability of
$\sim 95\%$. On the other hand,  $\eta\simeq 0.78$ comes out of the 
"histogram" MF estimate, which 
corresponds to reality probability of $\sim 56\%$. Then we can conclude
that this dip of MF seems to be statistically significant.\\

Finally, it is possible to obtain an estimate of the cluster mass by
integrating the luminosity function. In detail,

\begin{equation}
\label{mass}
M=<m>\times N=N\int \limits_{M_{V_{min}}}^{M_{V_{max}}} m(M_V)\phi (M_V)dM_V
\end{equation}

where $M_V$ is absolute stellar magnitude $M_V=V-(m-M)_V$, $\phi(M_V)$
is the normalized luminosity function

\begin{equation}
\label{phi}
\phi(M_V)=f(M_V)/N ,
\end{equation}

N is the total number of stars, $m(M_V)$ is the adopted mass-luminosity
relation. The luminosity function and mass-luminosity relation have
been approximated by a spline. Limits in the integral are $M_{V_{min}}=-0.1$
and $M_{V_{max}}=9.9$. The integral turns out to be
$$M\simeq 1620\pm 690 M_{\odot}$$
for kernel LF estimate and
$$M\simeq 1820\pm 650 M_{\odot}$$
for histogram LF estimate.

This mass estimate refers to total central field. Unfortunately it is very
difficult to evaluate what part of total cluster mass is constituting by
this estimate. It depends  on both unknown cluster size and unknown
upper limit of magnitude. If we assume that NGC~4815 has a limiting 
magnitude of  $V_c=20.8^m$, we get the mass estimate
$$M\simeq 880\pm 230 M_{\odot}$$.

\noindent
The error is determined from mass estimate for upper and lower boundaries
of confidence interval for LF (see Fig.5).

\section{Discussion and conclusions}
The MF can not be represented by a unique power law. In the mass range
$2.5 < m/m_{\odot}<0.8$,
the PDMF is reproduced by a power-law with $\alpha=3.1\pm0.3$ ($\alpha=2.35$
for the Salpeter MF in this notation).
In the same mass range the 
slope of the MF from the $I$ LF is $\alpha=2.9\pm0.3$. \\
Below $0.8 m_{\odot}$, the MF is much flatter
($\alpha = -0.68\pm0.49$).  \\
It is difficult to correctly interpret the PDMF in this 
low mass regime.
However we believe that the flattening of the PDMF below 
$0.8 m_{\odot}$ can be due to basically two effects.\\
First of all, we have already noted how the cluster vanishes (but not
disappears) into the Galactic field below $V = 21$, i.e. $m = 0.8
m_{\odot}$. 
This is likely originated  by irregularities in
the stellar background. As a consequence
low mass stars are preferentially missed 
due to absorption.\\
Secondly, we cannot however exclude cluster
dynamical evolution, which causes a loss of preferentially low mass
stars. Also, stripping of low mass stars
can occur along the cluster orbit,
due to the interaction
of the cluster with the dense Galactic field. NGC~4815 is indeed located
close to the Sagittarius-Carina spiral arm, rich in HII regions
and molecular clouds.
Estimating the relative importance of these two effects is
very difficult and goes beyond the aim of this paper.\\
Therefore, the slope of the MF below 0.8 $m_{\odot}$ can not be
considered representative of the cluster IMF.\-
 
On the other hand, whether 
the PDMF above 0.8 $m_{\odot}$ is
representative of the IMF is a matter which needs a careful model of
the cluster too.\\
A recent review on the IMF derived from the star counts is in Scalo
\cite*{scal98}. From the comparison of several studies, he found that
the spread in the slope at all masses above about $1m_{\odot}$ is too
large for adopting some average value, and in particular, there is  no
evidence for a clustering of points around the Salpeter value
(2.35). Scalo \cite*{scal98} concluded that either the systematic
uncertainties are so large that the IMF cannot yet be estimated, or
that there are real and significant variations of the IMF slope.  For
sure, the dynamical evolution, which differs from cluster to cluster,
can modify the MF slope, and also this effect contributes to the MF
slope dispersion.  It has emerged that the MF can not be
reproduced with a single power low: at low-masses the IMF from the
field stars, open and globular clusters \cite*{piot99} 
is relatively flat, and it becomes steeper above about
$1m_{\odot}$. Also NGC~4815 seems to show this break in the MF shape
at $m\sim 1m_\odot$ (Fig.~\ref{fmv}), though it is not possible to
disentangle the real differences in the IMF from the effects of the
dynamical evolution.
A final point is the effect of unresolved binaries and mass segregation.
As discussed by Scalo (1998, and references therein), 
unresolved binaries and mass segregation can only flatten the apparent 
IMF derived by not accounting for these, so the real IMF must be steeper 
than the quoted slope by an unknown amount.

In Fig.~\ref{allfm}, the MF of NGC 4815 is compared with the MFs for other
15 open clusters. Most of them cover masses above
$1m_{\odot}$. 
We have derived the MFs directly from the published LFs using the same
set of isochrones from Girardi et al. \cite*{gira00} used to get the
NGC~4815 MF. The MFs in Fig.~\ref{allfm} are shifted by arbitrary
constants for display purposes. The MF of NGC~4815 is one of the most
extended (in mass) MFs obtained so far for an open cluster.  Only for
the better known and nearby open clusters (Pleiades and M35), and for
objects observed with HST (NGC 2516, NGC 2477, and NGC 2420) do the MFs
reach $\simeq$ 0.1 $m_{\odot}$.

\begin{figure}[!h]
\centerline{\psfig{file=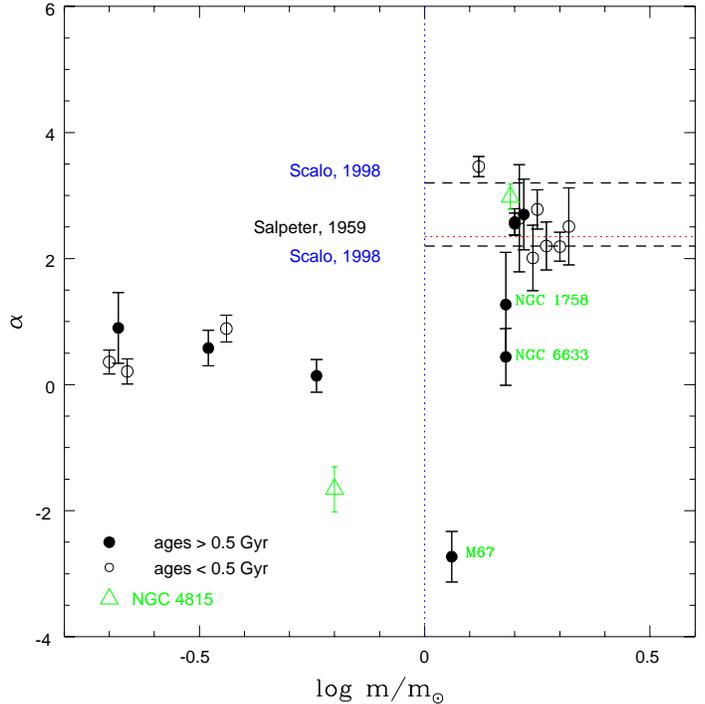,width=10cm,height=10cm}}
\caption{The MF slopes $\alpha$ are plotted as a function of the mean
fitting mass in the two mass ranges: $m>1m_{\odot}$ and
$m<1m_{\odot}$. Filled circles represent clusters older than 
0.5  Gyr, whereas open cirles stand for younger clusters. Note
the similarity of the MF slopes for the youngest clusters.}
\label{alfaeta}
\end{figure}

Fig. \ref{alfaeta} shows the slopes $\alpha$ of the MFs plotted in
Fig.~\ref{allfm} as a function of the mean mass. The slopes have
been obtained by fitting a power law in the $m<1m_\odot$ and
$m>1m_\odot$ mass ranges. Clusters with ages greater and less than or equal
to 0.5 Gyr are plotted with different symbols.
As noted by Scalo, the spread in slope is very large. However, it is
interesting to note that {\it all} the clusters with ages $\leq0.5$
Gyr have very similar slopes ($\alpha=2.25\pm0.88$) for $m>1m_\odot$.

NGC~4815, which has an age of 0.5 Gyr (Carraro and Ortolani
\cite*{carr94}), has a MF slope fully consistent with this value.  The
MFs of the clusters older than 0.5~Gyr have a similar or flatter slope. This
may indicate that their PDMF may have not or may have been changed by
the dynamical evolution which has had more time to operate. A
dynamical model of each of these clusters is necessary to confirm this 
interpretation of Fig.~\ref{alfaeta}.
At the moment, the similarity of the MF slopes for
all the youngest clusters in Fig.~\ref{alfaeta} can be interpreted as
an evidence of a similarity of the IMF, which seems to not depend much
on metallicity, stellar density, or galactocentric radius.

The interpretation of the MF slopes below $0.8 m_\odot$ is more
difficult, both because there are less (and more uncertain) data, and
because we expect that in this mass regime the dynamical evolution
should be more effective in altering the IMF. Still, all the MFs seems
to have the same flat general trend. We also note that these slopes
represent an upper limit for the slopes of the MF of the Galactic
globular clusters (Piotto and Zoccali \cite*{piot99}).

\begin{acknowledgements}
We acknowledge the financial support by the Ministero della Ricerca
Scientifica e Tecnologica (MURST) under the program {Treatment of large field
astronomical images}. PBS an AFS gratefully acknowledge
the generosity of the Universit\'a di Padova and MURST for supporting
their visit to the Dipartimento di Astronomia and the Osservatorio
di Padova.
\end{acknowledgements}
\bibliographystyle{aabib99}
\bibliography{mnemonic,biblio}
\end{document}